\documentclass{article}
\usepackage{spconf,amsmath,graphicx,hyperref, booktabs, makecell, multirow, caption, booktabs, subcaption, comment, xcolor}
\usepackage{amssymb, multirow}
\usepackage{pifont}
\usepackage{soul} 
\newcommand{\cmark}{\ding{51}}%
\newcommand{\xmark}{\ding{55}}%
\newcommand{\fcriptsize}{\fontsize{7.3pt}{9pt}\selectfont}

\title{When Do Self-Supervised Representations Help Speech Restoration?}

\name{Yongjoon Lee$^{1}$ \qquad Younghoo Kwon$^{1}$ \qquad Jung-Woo Choi$^{2 *}$\thanks{$^*$Corresponding author.}}
\address{$^{1}$School of Electrical Engineering, KAIST, Daejeon, South Korea \\
$^{2}$Department of AI Systems, KAIST, Daejeon, South Korea}
      
\begin{document}
\ninept
\maketitle
\begin{abstract}

Self-supervised learning (SSL) representations are widely used in speech restoration (SR), and the prevailing trend is SSL feature cleansing: fine-tuning the SSL to recover clean features from degraded speech, then conditioning the restoration model on them. However, the representational capability of SSL when encoding degraded speech has been underexplored, and we ask whether cleansing always benefits SR training. Across SSL encoders and perceptual quality proxies, we discover perceptual correlation collapse---under severe degradation, cleansing degraded SSL features toward clean ones misaligns with the SR objective. We hypothesize the restoration backbone can supply the SSL with clues to resolve this collapse, since its intermediate layers already produce partially restored features. We therefore propose Early Restoration Injection (ERI), which routes intermediate features from the restoration backbone into the SSL through a lightweight conditioning network. ERI excels over the unconditioned baseline and, on the VoiceFixer benchmark, achieves competitive performance among multiple baselines.

\end{abstract}

\vspace{-0.5em}
\section{Introduction}
\label{sec:intro}
\vspace{-0.5em}
Speech restoration (SR) aims to recover clean speech from degraded speech, where the degradation may be noise, reverberation, bandwidth limitation, clipping.  Self-supervised learning (SSL) representations~\cite{Chen_2022, baevski2020wav2vec20frameworkselfsupervised} have been widely used to address SR for their rich acoustic and phonetic information, and strong alignment with perceptual quality~\cite{babaev2024finallyfastuniversalspeech}. One line of research feeds both the SR output and the clean speech into an SSL separately and minimizes the loss between the two resulting representations, utilizing SSL as a guiding signal in training. MaskSR2~\cite{liu2024jointsemanticknowledgedistillation} uses HuBERT~\cite{hsu2021hubertselfsupervisedspeechrepresentation} as a semantic teacher to conduct knowledge distillation, improving intelligibility. Universe++~\cite{scheibler2024universalscorebasedspeechenhancement} uses an SSL-finetuned phoneme prediction model~\cite{xu2021simpleeffectivezeroshotcrosslingual} to minimize the connectionist temporal classification loss~\cite{10.1145/1143844.1143891} between the predicted phonemes of the clean and restored speech. In this way, rich SSL representations can be efficiently learned since SSL is only required in the training phase. 


Another line of research feeds the degraded speech directly to speech SSL and then injects the resulting degraded SSL representation into the SR model as a conditioning feature. Early efforts~\cite{song2022exploringwavlmspeechenhancement, Byun_2023} utilize WavLM~\cite{Chen_2022} as a feature encoder for speech enhancement, paving the way for SSL-conditioned SR. FINALLY~\cite{babaev2024finallyfastuniversalspeech} uses WavLM features as conditioning signals to jointly perform speech enhancement and bandwidth extension. These methods directly integrate SSL models into the enhancement model's forward path, effectively leveraging the rich representational power of SSLs.


Beyond using SSL representations as a mere guiding signal in training or as conditioning features, recent approaches first restore the SSL features, then use cleansed features as conditions for SR models. 
This so-called \textbf{SSL feature cleansing} recovers clean SSL features given a degraded speech waveform either by using language models (LM) or by fine-tuning SSL itself. GenSE~\cite{yao2025gensegenerativespeechenhancement} and SenSE~\cite{li2026sensesemanticawarehighfidelityuniversal} are two representative methods that use LM to generate the recovered SSL features given noisy SSL features. Miipher~\cite{koizumi2023miipherrobustspeechrestoration}, SIDON~\cite{nakata2026sidonfastrobustopensource}, PASE~\cite{rong2025paseleveragingphonologicalprior}, and StuPASE~\cite{rong2026stupaselowhallucinationstudioqualitygenerative} incorporate pretraining where SSL models are fine-tuned to restore the clean SSL features by themselves. LM-based SSL feature cleansing requires demanding computational resources, while fine-tuning-based SSL feature cleansing is relatively less compute- and memory-intensive. 


Though fine-tuning SSL for feature cleansing has proven effective for SR tasks, existing approaches have mainly depended on empirical results to show these performance gains. SSL feature distance has been studied as a property of the representation — how a layer's clean–degraded distance relates to its importance for speech enhancement~\cite{hung2022boostingselfsupervisedembeddingsspeech}. Feature cleansing, by contrast, minimizes this distance, and in doing so presupposes that a smaller distance implies better restored speech — an assumption that has not been examined. We pose a fundamental question: \emph{is SSL feature cleansing always effective for SR?} Across multiple SSL encoders and perceptual quality proxies, we find \textbf{perceptual correlation collapse}: the correlation between SSL feature distance and perceptual quality holds under mild degradation but breaks past a critical severity, so that minimizing the SSL loss no longer tracks quality beyond this point.



One simple solution to this phenomenon is to place a preliminary restoration model before the SSL so that the SSL \emph{never sees} severely degraded speech and instead consumes the decoded waveform that model produces; this buys clean-enough input, but at the cost of an auxiliary model. 
In fact, the SR backbone we already have produces partially restored features at its intermediate layers. This leads to our central hypothesis: if those features are already restored enough to serve as evidence, exposing them to the SSL should mitigate perceptual correlation collapse, obtaining the benefit of pre-restoration without a separate pre-restoration model. To validate this, we propose \textbf{Early Restoration Injection (ERI)}, a method that injects intermediate layer SR features into the SSL via a simple conditioning network. This method effectively resolves the perceptual correlation collapse by feeding SSL with partially recovered speech features, converging to higher quality than the unconditioned baseline at a matched parameter count. Our contributions are two-fold:

\begin{itemize}

\item We discover \textbf{perceptual correlation collapse}: as degradation increases, cleansing degraded SSL features does not fully track perceptual quality. We establish this with analysis across multiple SSL models and perceptual-quality proxies.

\item We propose \textbf{Early Restoration Injection (ERI)}, a simple yet novel method that injects intermediate features from the SR network into SSL feature cleansing. Our framework achieves competitive performance on the VoiceFixer benchmark.
    
\end{itemize}

\begin{figure}[!t]
  \centering
    \includegraphics[width=0.8\linewidth]{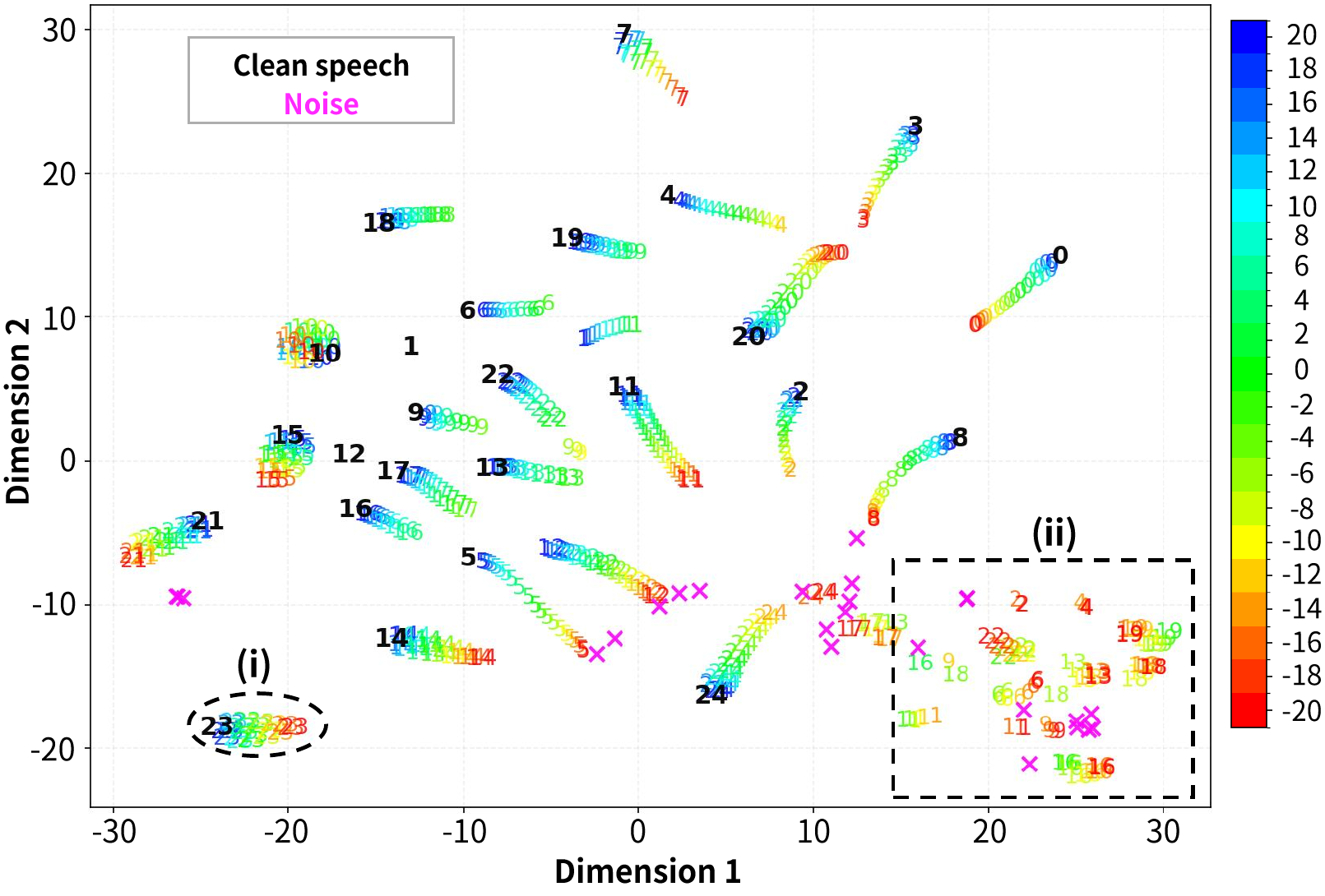}
  \caption{t-SNE plots of speech degraded with additive noise at varying SNR level, visualized in WavLM feature space.}
  \label{fig:tsne_combined}
\end{figure}

\begin{figure}[t]
    \centering
    \begin{subfigure}[b]{\linewidth}
        \centering
        \includegraphics[width=\linewidth]{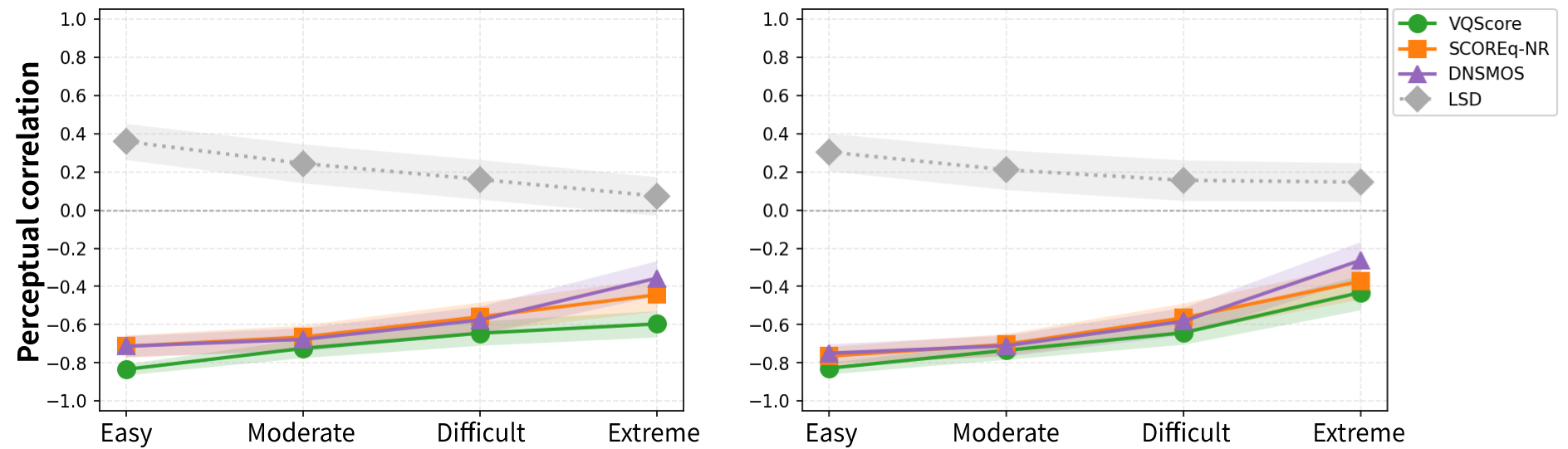}
        \caption{WavLM (Left) and HuBERT (Right)}
        \label{fig:pcc_wavlm}
    \end{subfigure}\vfill
    \begin{subfigure}[b]{\linewidth}
        \centering
        \includegraphics[width=\linewidth]{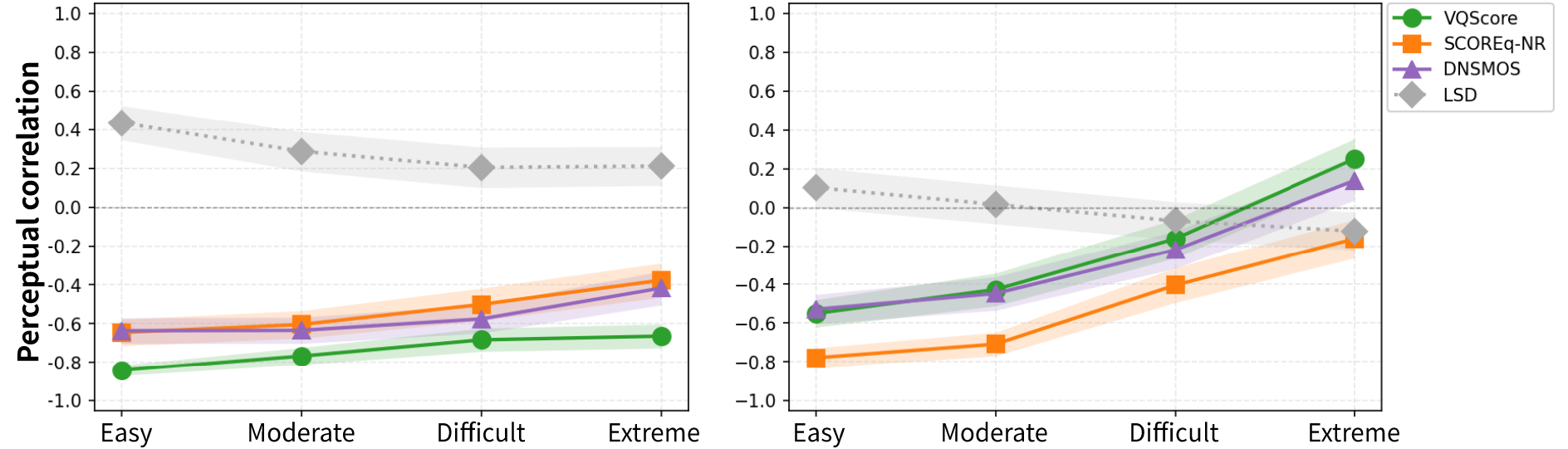}
        \caption{SPEAR (Left) and Wav2Vec2 (Right)}
        \label{fig:pcc_hubert}
    \end{subfigure}
    \vspace{-1.0em}
    \caption{SSL-wise plot of perceptual correlation.}
    \label{fig:pcc_all}
\end{figure}

\section{When do self-supervised representations help speech restoration?}
\label{sec:analysis}
\vspace{-0.5em}
We first show how SSL represents degraded speech, using a motivating visualization in which some SSL features lose the speech identity and become indistinguishable from those of noise even when the speech is audible. Then we discover \textbf{perceptual correlation collapse}, a phenomenon where SSL feature cleansing does not correspond to perceptual quality increase if speech degradation is severe.


\subsection{Some utterances lose their identity in the SSL feature space at low SNRs, while others remain firmly anchored to their clean counterparts.}
\label{sec:tsne}
\vspace{-0.3em}

\textbf{Method} \noindent We design a toy visualization using t-SNE~\cite{vanDerMaaten2008}. We sample $m$ speech utterances from FLEURS-R~\cite{ma2024fleursrrestoredmultilingualspeech} and LibriTTS-R~\cite{koizumi2023librittsrrestoredmultispeakertexttospeech} and $m$ noise samples from WHAM!~\cite{Wichern2019WHAM} and DNS Challenge 2020~\cite{reddy2020interspeech2020deepnoise}, thus formulating a pair $(x_j,n_j)$, where $x$, $n$, and $j$ represent speech data, noise data, and the data index, respectively. For each pair, we synthesize noisy copies at SNR ranging from $-20$ to $20$\,dB in $2$\,dB increments. Together with the clean speech and the noise data, each data pair of index $j$ yields 23 conditions. The samples are then fed individually to the SSL, and the resulting feature representations are visualized using t-SNE. We used the final-layer feature (time frame-pooled) of WavLM~\cite{Chen_2022} for the visualization. We set $m$ to 25.

\noindent\textbf{Finding} Fig.~\ref{fig:tsne_combined} shows t-SNE embeddings of WavLM features for a set of utterances, each mixed with its own noise sample across several SNR levels; the utterance index $j$ is marked at each point, with clean speech in bold black and noise in pink. Since SNR is the only source of variation within an utterance, a representation encoding speech content should organize by index, with degraded samples remaining near their clean anchor. Region~(i) behaves this way: samples sharing an index stay together, and SNR governs how far each sits from its anchor. Region~(ii) does not. Degraded samples do not simply disperse; they converge onto the region occupied by the noise embeddings, despite each utterance carrying a different noise sample. Most importantly, they do not arrive together: some noisy samples reach region (ii) at around $0$\,dB (See $j=11,16,19$), where speech and noise carry similar power, while others at the same SNR stay beside their anchors. We can see that some utterances lose their speech features even when the speech is audible. 

\subsection{Lower SSL loss does not always translate into higher perceptual quality, especially when speech degradation is severe.}
\label{sec:analysis2}
\vspace{-0.3em}

From the t-SNE visualization, we observe that some utterances lose their utterance identity in SSL feature space when they are not supposed to (e.g., 0 dB; even when the speech is audible). Thus, we conjecture that in multi-degradation scenarios, the SSL feature distance would not reflect the true perceptual quality if the degradation severity is beyond a certain level. However, a single degradation configuration (e.g., SNR) cannot represent the overall perceptual quality in a multi-degradation scenario, motivating us to use perceptual quality evaluation metrics as proxies for thorough analysis.

\noindent \textbf{Method} \noindent We randomly sample $m$ clean English utterances $x_1$, $x_2$, $\ldots$,$x_m$ from LibriTTS-R~\cite{koizumi2023librittsrrestoredmultispeakertexttospeech}. For each clean utterance $x_j$ where $j=1,...,m$, four degraded versions $y_j^{(E)},y_j^{(M)}, y_j^{(D)},y_j^{(X)}$, corresponding to Easy ($E$), Moderate ($M$), Difficult ($D$), and Extreme ($X$), respectively, are simulated. The degradation type and the application probability are homogeneous across different degradation severities. The degradation pipeline uses noise, bandwidth limitation, clipping, reverb, and packet loss. Each degradation is sampled from a severity-specific range, chosen to be harsher at every step from Easy to Extreme. Noise SNR drops from $[12.5, 20]$ to $[-10, -2.5]$ dB, the bandwidth cutoff from $[5750, 7000]$ to $[2000, 3250]$ Hz, clipping thresholds tighten inward from $[0.00, 0.02] / [0.98, 1.00]$ to $[0.06, 0.08]/[0.92, 0.94]$, and packet loss duration grows from 5 to 20 ms.


For each degradation severity, we compute the cosine distance $d_j^{(E)}, d_j^{(M)}, d_j^{(D)}, d_j^{(X)}$ between the SSL output features for clean and degraded speech. The perceptual quality prediction of the degraded speech $q_j^{(E)}, q_j^{(M)}, q_j^{(D)},q_j^{(X)}$ is computed. For a general analysis, we select multiple SSLs and quality prediction models. For SSL, we use WavLM~\cite{Chen_2022}, HuBERT~\cite{hsu2021hubertselfsupervisedspeechrepresentation}, SPEAR~\cite{yang2026spearunifiedsslframework}, and Wav2Vec 2.0~\cite{baevski2020wav2vec20frameworkselfsupervised}, all in \texttt{large} variants. For neural quality prediction, we use VQScore~\cite{fu2024selfsupervisedspeechqualityestimation}, SCOREQ~\cite{ragano2025scoreqspeechqualityassessment}, and DNSMOS~\cite{reddy2021dnsmosnonintrusiveperceptualobjective}. We use the no-reference (NR)  version of SCOREQ. Finally, we use log spectral distance (LSD) for model-free quality estimation. To quantify the correlation between the SSL distance and quality score, we compute the Spearman correlation $\rho^{(\mathrm{sev})}$~\cite{spearman04} over the $m$ paired observations 
$ \{ (d_j,q_j) \}^m_{j=1}$, for each severity. $\mathrm{sev}$ represents degradation severity index, and the correlation coefficient represents \textbf{Perceptual correlation} of SSL loss. For robust analysis, we set $m$ to 100, and conduct bootstrapping for 1{,}000 times and plot both the mean correlation and standard deviation. The SSL output feature is computed as the time frame-wise mean of the final layer output.  

\noindent \textbf{Finding} \noindent Fig.~\ref{fig:pcc_all} shows that the perceptual correlation (SSL-loss/quality correlation) is strong under mild degradation but collapses beyond a critical severity, implying SSL feature distance between clean and severely degraded speech does not rank-order perceptual quality well. This has a consequence for SSL feature cleansing: minimizing SSL loss does not always translate into higher perceptual quality when the speech degradation is severe. 

We verify that this collapse is not an artifact of any single design choice in the analysis. First, it is not specific to a particular SSL model: the same trend appears across all four representations, indicating that the collapse reflects a property of the SSL loss-quality relationship rather than a trait of one SSL feature space. Second, to rule out sampling noise as the cause, we bootstrap each correlation estimate and report the resulting standard deviation in Fig.~\ref{fig:pcc_all}; the collapse persists well outside this uncertainty band, confirming that it is a stable trend rather than a sampling artifact. Third, to check whether the collapse is instead an artifact of the perceptual quality metric itself, we repeat the analysis with LSD, a model-free quality estimator computed directly from the log-spectral distance. Although LSD may not perfectly track subjective perceptual quality, it exhibits the same collapse beyond the critical severity, which suggests the effect is a genuine property of the SSL representation rather than an artifact of the particular quality metric used to measure it. 

Together, these results expose a critical problem for SSL-conditioned SR. 
An SSL distance objective provides less useful gradient toward perceptual quality in the severe regime that restoration must handle. We therefore explain our proposed method to provide the SSL encoder with restored features in the next section.
\vspace{-0.5em}
\section{Early Restoration Injection}
\begin{figure}[t]
\centering
\includegraphics[width=0.32\textwidth]{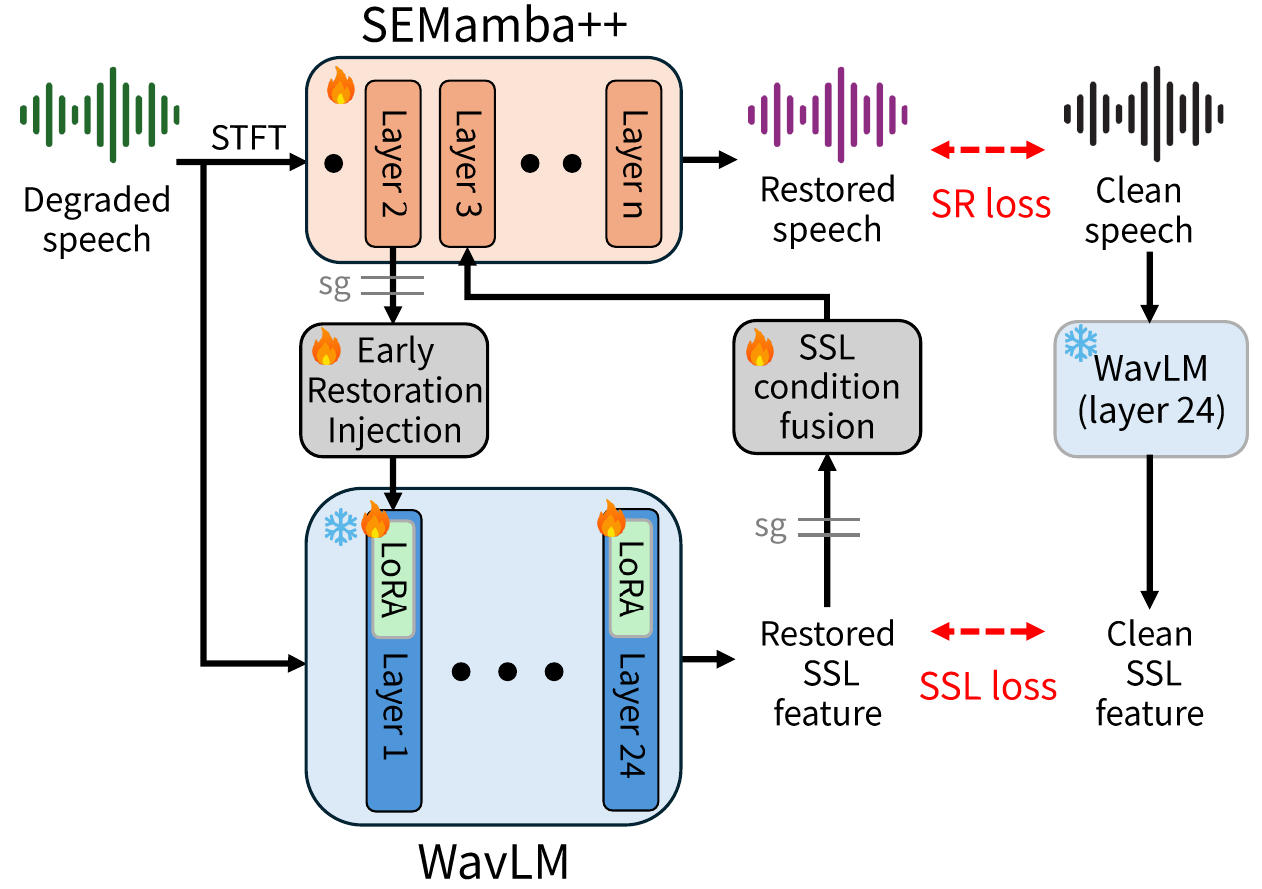}
\vspace{-0.5em}
\caption{Model architecture and training framework.}
\label{fig:modelarch}
\end{figure}

\vspace{-0.5em}
Fig.~\ref{fig:modelarch} shows the overall framework of ERI. SEMamba++ \cite{lee2026semambageneralspeechrestoration} and WavLM~\cite{Chen_2022} are used as the SR and speech SSL models, respectively. SEMamba++ comprises an encoder, bottleneck, and decoder. The encoder encodes the degraded speech into hidden features. The bottleneck is a 4-layer sequence of time-frequency processing modules. The decoder maps the restored features into the magnitude and phase spectrogram, converting to the clean waveform. WavLM comprises a feature extractor followed by 24 transformer layers. 

ERI is a simple method to feed the SSL with the latent bottleneck feature of SEMamba++. The latent bottleneck feature has a shape of $T, F, D$, where $T$, $F$, and $D$ represent the time, frequency, and channel dimensions of SEMamba++. ERI first reshapes the latent bottleneck feature to $T, FD$, followed by a linear layer that maps the dimension $FD$ to $C$, where $C$ represents the channel dimension of WavLM. After applying Layer normalization and LeakyReLU activation, we apply adaptive average pooling to match the time dimension to $T'$, corresponding to the time dimension of WavLM. The resulting feature is concatenated with the WavLM feature of shape $T', C$ after the convolutional feature extractor, deriving the feature of shape $T', 2C$. Then a linear layer maps the feature into $T', C$ so that the condition injection is made.

\label{sec:otherdetail}
WavLM is fine-tuned with Low-Rank Adaptation (LoRA)~\cite{hu2021loralowrankadaptationlarge}, where the hyperparameters $r$ and $\alpha$ were set to 16 and 32, respectively. After WavLM takes the degraded waveform and the ERI condition to restore the SSL feature, the restored SSL feature is injected back into the SR backbone via the SSL condition fusion module. The SSL condition fusion module is an adaptive layer normalization (AdaLN)~\cite{xu2019understandingimprovinglayernormalization}-based module fusing the SSL condition into SEMamba++. We use the layer 24 output from WavLM to conduct SSL condition fusion, as done by previous work~\cite{rong2025paseleveragingphonologicalprior}. We extract the ERI condition from SEMamba++ layer 2 and fuse the cleansed SSL features back at layer 3. 

We apply stop-gradient (sg) before ERI and SSL condition fusion. sg before ERI blocks the SSL loss from reaching SEMamba++. In particular, this prevents the SSL loss from acting as a representation alignment~\cite{yu2025representationalignmentgenerationtraining} loss: SEMamba++ is trained only by the SR objective, so any effect of ERI is gained via the forward path rather than by SSL gradients shaping SEMamba++. sg before SSL condition fusion prevents the spectrogram-level SR loss from reaching WavLM, keeping it focused on the SSL feature cleansing. We found empirically that without sg before SSL condition fusion, the SSL loss was not effectively minimized and performance degraded.

\vspace{-1.0em}
\subsection{Training objective}
\vspace{-0.5em}
The training objective can be divided into learning SR and SSL models. For training the SR model, we followed the LSGAN~\cite{mao2017squaresgenerativeadversarialnetworks} configuration from SEMamba++. For training the SSL feature cleansing, we utilized an L2-normalized MSE loss of the 24th-layer features $\mathcal{L}_{\mathrm{SSL}}$. We append $\mathcal{L}_{\mathrm{SSL}}$ to the generator loss so that the training can be conducted end-to-end, simultaneously optimizing both the SR and SSL models. The final generator loss is  $\mathcal{L}_{\mathrm{total}}=\mathcal{L}_G+\mathcal{L}_{\mathrm{SSL}}$. 


\vspace{-0.5em}
\section{Experimental settings}
\label{ssec:experiments}
\vspace{-0.5em}

\subsection{Training}
\label{sssec:training}
\vspace{-0.5em}

\textbf{Datasets} \label{sec:traindata}For speech, we used VCTK-0.92~\cite{vctk} and LibriTTS-R~\cite{koizumi2023librittsrrestoredmultispeakertexttospeech}. For VCTK, we excluded speakers in p280 and p315 following~\cite{Liu_2022}. For LibriTTS-R, we used the train-clean-100 and -360 subsets, totaling ~300 hours of speech sources. For noise data, we used WHAM!~\cite{Wichern2019WHAM} and DNS challenge~\cite{reddy2020interspeech2020deepnoise}. For RIR, we used Arni~\cite{prawda2022calibrating} and the DNS challenge~\cite{reddy2020interspeech2020deepnoise}. All data were sampled at 16 kHz and split into a train and a testset, including the noise and RIR data. Note that LibriTTS-R already has a predetermined testset.

\noindent\textbf{Degradation simulation} Noise, reverberation, bandwidth limitation, and clipping were used to simulate degradation. In training, various distortions are randomly applied with predetermined probabilities. The signal-to-noise ratio (SNR) ranged from -5 to 20 dB, and bandwidth reduction was applied at various cutoff frequencies (2-7 kHz). For clipping, we followed the official configurations~\cite{URGENT-Zhang2024}.

\noindent\textbf{Training configurations} The segment size of 48,000 samples (3 seconds) was used for training. For STFT, we used a Hann window with a window length of 400 samples and a hop size of 160 samples. We set the learning rate to 1e-4 and used the AdamW optimizer with $\beta_{1}$ and $\beta_{2}$ set to 0.8 and 0.99, respectively. We set the batch size to 8 and used exponential learning rate decay with $\gamma=0.99$. ERI was trained in two stages. In stage 1, SEMamba++ was trained with the same data configuration. In stage 2, we trained the full ERI framework for an additional 300k steps. 

\vspace{-0.5em}

\subsection{Evaluation}
\label{sec:evaluation}
\vspace{-0.5em}
\textbf{Datasets} For comparisons within our model, we used a simulated set built from the test split in Section~\ref{sec:traindata}. For system comparison against other methods, we used the official VoiceFixer test set~\cite{Liu_2022}, which covers noise, reverb, bandwidth limitation, and clipping.

\noindent\textbf{Baseline methods} For GAN methods, we used VoiceFixer~\cite{Liu_2022} and SEMamba++~\cite{lee2026semambageneralspeechrestoration}. For generative methods, we used AnyEnhance~\cite{Zhang_2025}. For SSL-based methods, we used SIDON~\cite{nakata2026sidonfastrobustopensource}. Lastly, for regressive methods, we used TF-GridNet~\cite{wang2023tfgridnetmakingtimefrequencydomain}. For all models, we used the official checkpoint.


\begin{table}[t]
\centering
\scriptsize
\setlength{\tabcolsep}{3pt}
\caption{Performance evaluation on the VoiceFixer testset.}
\vspace{-0.7em}
\renewcommand{\arraystretch}{0.92}
\begin{tabular}{lcccccc}
\toprule
Model & \makecell{\# Params (M) \\ total / trainable} & OVRL$^\uparrow$ & UTMOS$^\uparrow$ & PESQ$^\uparrow$ & ESTOI$^\uparrow$ & LPS$^\uparrow$ \\
\midrule
Degraded  & -     & 2.39 & 2.63 & 1.94 & 0.58 & 0.84 \\
\midrule
VoiceFixer  & 112 / 112 & 3.00 & 3.44 & 2.04 & 0.56 & 0.80 \\
TF-GridNet & 14.4 / 14.4  & 3.06 & 3.41 & 2.47 & 0.68 & \textbf{0.88} \\
AnyEnhance   & 364 / 324 & \underline{3.14} & 3.75 & 2.45 & 0.67 & \textbf{0.88} \\
SIDON   & 250 / 57.4 & \textbf{3.20} & \textbf{3.97} & 2.05 & 0.58 & \textbf{0.88} \\
SEMamba++& 2.7 / 2.7 & 3.10 & 3.75 & \underline{2.76} & \underline{0.70} & 0.87 \\ 
\midrule
ERI (Ours) & 340 / 24.5  & 3.10 & \underline{3.86} & \textbf{2.82} & \textbf{0.71} & \textbf{0.92} \\
\bottomrule
\end{tabular}
\label{tab:voicefixerresults}
\end{table}

\noindent\textbf{Evaluation Metrics} 
We used SCOREQ~\cite{ragano2025scoreqspeechqualityassessment}, UTMOS~\cite{saeki2022utmosutokyosarulabvoicemoschallenge}, and OVRL~\cite{reddy2021dnsmosnonintrusiveperceptualobjective} to assess speech perceptual quality. For SCOREQ, we used the no-reference mode with a synthetic data setup. To evaluate signal fidelity, we used Perceptual Evaluation of Speech Quality (PESQ)~\cite{941023}. We used extended short-time objective intelligibility (ESTOI)~\cite{7539284} to measure intelligibility. Lastly, Levenshtein phoneme similarity (LPS)~\cite{10363040} was used to capture phoneme similarity, which is useful in evaluating generative SR models. 

\vspace{-0.2em}
\section{Experimental results}
\vspace{-0.5em}

Table~\ref{tab:voicefixerresults} illustrates the results of multiple baseline models evaluated on the VoiceFixer testset for benchmarking SR. ERI attains the best PESQ, ESTOI, and LPS, indicating strong signal fidelity and intelligibility in the restored speech, while its second-best UTMOS keeps perceptual quality competitive. Although ERI substantially increases the parameter count due to the added SSL and conditioning modules, its real-time factor (measured on a single RTX 4090 GPU) increases only from 0.015 for SEMamba++ to 0.020 for ERI, showing that the inference efficiency is not heavily compromised.


Table~\ref{tab:allperf} measures the effectiveness of ERI (\cmark) compared to the baseline (\xmark). The baseline minimizes SSL loss without ERI features. To compensate for the parameter increase in ERI, we double the hyperparameters $r$ and $\alpha$ and attach LoRA to both feed-forward networks (FFNs) rather than the second FFN alone (i.e., default design), resulting in a parameter gap of less than 1M. ERI improves performance over the baseline, showing the effectiveness of intermediate features from the SR backbone for SSL feature cleansing. Moreover, since ERI feeds restored intermediate features of the SR backbone, the effectiveness of ERI could be compromised if the SR backbone is not fully converged. To test this, we instantiate ERI with the SR backbone trained less in Stage 1 (i.e., 300k steps). As shown in Table~\ref{tab:allperf}, ERI remains effective even with such a setup, showing that ERI does not require full convergence of the SR backbone for its success. Lastly, to show that our method could be applied to other SSLs analyzed in Section~\ref{sec:analysis}, we design an ERI variant where the SSL is HuBERT. The result shows that the ERI remains its strengths for different SSL as well, implying the general applicability.

\begin{table}[t]
\centering
\scriptsize
\caption{Performance comparison on ERI.}
\vspace{-0.7em}
\renewcommand{\arraystretch}{0.92}
\setlength{\tabcolsep}{4pt}
\begin{tabular}{ccc|cccc}
\toprule
ERI & SSL & Stage 1 steps (k)  & $\text{SCOREQ}^{\uparrow}$ & $\text{UTMOS}^{\uparrow}$ & $\text{PESQ}^{\uparrow}$ & $\text{LPS}^{\uparrow}$ \\
\midrule

 \cmark & \multirow{2}{*}{WavLM} & \multirow{2}{*}{500 (Full)}
& \textbf{3.97}
& \textbf{4.06}
& \textbf{2.50}
& \textbf{0.90}\\

\xmark &  & 
& 3.93
& 4.03
& 2.42
& 0.88\\
\midrule


\cmark  & \multirow{2}{*}{WavLM} & \multirow{2}{*}{300}
& \textbf{3.95}
& \textbf{4.04}
& \textbf{2.41}
&  \textbf{0.89}\\

\xmark  &  & 
& 3.88
& 3.99
& \textbf{2.41}
& 0.88\\

\midrule
\cmark  & \multirow{2}{*}{HuBERT} & \multirow{2}{*}{500}
& \textbf{3.93}
& \textbf{4.03}
& \textbf{2.41}
& \textbf{0.87}\\


\xmark  &  & 
& 3.86
& 3.96
& 2.33
& 0.83\\

\bottomrule
\end{tabular}
\label{tab:allperf}
\end{table}

\begin{table}[t]
\centering
\caption{Design study. We used 300k steps for stage 1 training for this study.}
\vspace{-0.7em}
\scriptsize
\renewcommand{\arraystretch}{0.9}
\setlength{\tabcolsep}{4.4pt}
\begin{tabular}{lc|cccc}
\toprule
Design   & Params & $\text{SCOREQ}^{\uparrow}$ & $\text{UTMOS}^{\uparrow}$ & $\text{PESQ}^{\uparrow}$ & $\text{LPS}^{\uparrow}$ \\
\midrule
Ours (default) & 24M &  \textbf{3.95}
& \textbf{4.04}
& 2.41
& \textbf{0.89}\\
ERI layer 2$\rightarrow$1 & 24M & 3.91
& 4.03
& \textbf{2.42}
& 0.88\\
w/o sg on SSL cond. fus. & 24M & 3.92 & 4.02 & \textbf{2.42} & 0.88\\

\midrule
\textcolor{gray}{AdaLN-based ERI} & \textcolor{gray}{32M} & \textcolor{gray}{3.94} & \textcolor{gray}{4.04} & \textcolor{gray}{2.43} & \textcolor{gray}{0.89}\\

\bottomrule
\end{tabular}
\label{tab:ablation}
\end{table}

\begin{figure}[ht!]
    \centering
    \includegraphics[width=0.76\linewidth]{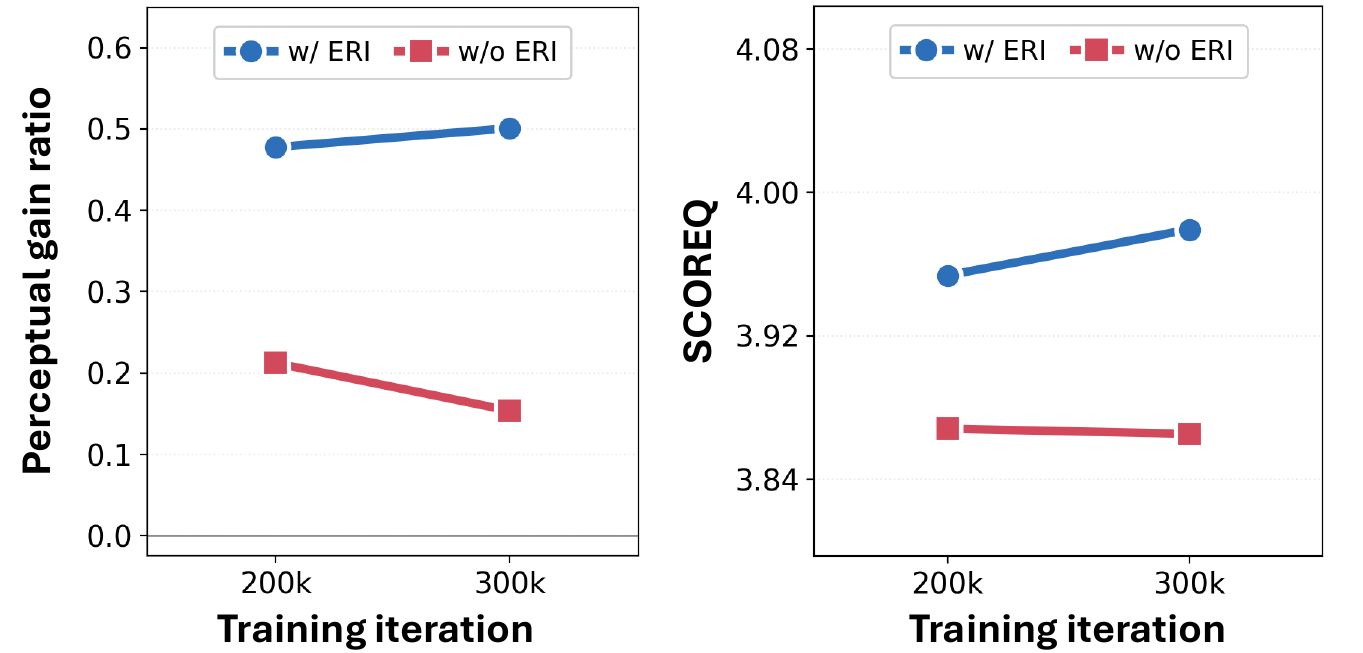}
    \vspace{-0.7em}
    \caption{Perceptual gain ratio and SCOREQ value.}    
    \label{fig:qgainratio}
\end{figure}

Fig.~\ref{fig:qgainratio} shows that ERI enables effective translation of SSL loss minimization into perceptual quality gain. Using SCOREQ and SSL loss values after initial training stabilization as anchors, we define \textbf{Perceptual gain ratio} as the SCOREQ increment per $10^{-3}$ reduction in SSL loss. The ratio substantially favors ERI, implying that SSL loss minimization more effectively leads to perceptual quality gain when equipped with ERI. Note that analyzing the perceptual correlation from Section~\ref{sec:analysis2} on a fully trained ERI is confounded, as at convergence both w/ ERI and w/o ERI learn to cleanse SSL features. The perceptual gain ratio instead directly measures whether ERI translates the SSL objective to perceptual quality in training.


\noindent \textbf{Design study} Table~\ref{tab:ablation} shows the design choice study results to measure the component-wise contribution in the proposed method. Modifying the ERI layer from 2 to 1 degrades perceptual quality, suggesting that injecting more restored features into the SSL backbone improves performance. Removing the sg on the SSL condition fusion degrades perceptual quality, attributable to the ineffective SSL loss minimization discussed in Section~\ref{sec:otherdetail}. Lastly, changing the ERI mechanism from simple concatenation (i.e., default design) to more complicated AdaLN did not result in a significant performance gain, implying that ERI's benefit comes from the intermediate-layer features themselves rather than the increased capacity.



\vspace{-0.3em}
\section{Conclusions}
\vspace{-0.5em}

We discover perceptual correlation collapse: under severe degradation, the SSL feature cleansing objective does not fully align with the restoration objective. 
Though our correlation analysis relies on neural quality proxies for human perception, the agreement we observe across multiple proxies shows the trend is not an artifact of a single metric. To address perceptual correlation collapse, we propose Early Restoration Injection (ERI), which conditions SSL feature cleansing on intermediate features from the restoration model, supplying the SSL with partially restored evidence in place of severely degraded input. ERI adds only simple operations, and our experiments attribute its gain to the injected features themselves rather than to added capacity or a more expressive fusion mechanism. ERI improves over the unconditioned baseline at matched parameter count, and performs competitively against multiple baseline models.


\fcriptsize{
\section{Acknowledgments}
\vspace{-1em}
This work was supported by the National Research Foundation of Korea (NRF) grant (No. RS-2024-00337945), the STEAM research grant (No. RS-2024-00464269) funded by the Ministry of Science and ICT of Korea government (MSIT), the BK21 FOUR program through the NRF grant funded by the Ministry of Education of Korea government (MOE), and Artificial intelligence industrial convergence cluster development project funded by the Ministry of Science and ICT (MSIT, Korea)\&Gwangju Metropolitan City.} 

\vspace{-1em}

\bibliographystyle{IEEEbib}
{\fcriptsize
\bibliography{strings,refs2}

@string{ECCV = "European Conference on Computer Vision"}

@string{NeurIPS = "Advances in Neural Information Processing Systems"}

@string{ICML = "International Conference on Machine Learning"}

@string{ICLR = "International Conference on Learning Representations"}

@string{AAAI = "Proceedings of the AAAI Conference on Artificial Intelligence"}

@string{ICASSP = "IEEE International Conference on Acoustics, Speech, and Signal Processing"}

@string{INTERSPEECH = "Interspeech"}

@string{ICME = "IEEE International Conference on Multimedia and Expo"}

@string{WASPAA = "Workshop on Applications of Signal Processing to Audio and Acoustics"}

@inproceedings{yu2025representationalignmentgenerationtraining,
      title={Representation Alignment for Generation: Training Diffusion Transformers Is Easier Than You Think}, 
      author={Sihyun Yu and Sangkyung Kwak and Huiwon Jang and Jongheon Jeong and Jonathan Huang and Jinwoo Shin and Saining Xie},
      year={2025},
      booktitle=ICLR,
      eprint={2410.06940},
      archivePrefix={arXiv},
      primaryClass={cs.CV},
      url={https://arxiv.org/abs/2410.06940}, 
}

@inproceedings{hung2022boostingselfsupervisedembeddingsspeech,
      title={Boosting Self-Supervised Embeddings for Speech Enhancement}, 
      author={Kuo-Hsuan Hung and Szu-wei Fu and Huan-Hsin Tseng and Hsin-Tien Chiang and Yu Tsao and Chii-Wann Lin},
      booktitle=INTERSPEECH,
      year={2022},
      eprint={2204.03339},
      archivePrefix={arXiv},
      primaryClass={eess.AS},
      url={https://arxiv.org/abs/2204.03339}, 
}

@inproceedings{koizumi2023miipherrobustspeechrestoration,
      title={Miipher: A Robust Speech Restoration Model Integrating Self-Supervised Speech and Text Representations}, 
      author={Yuma Koizumi and Heiga Zen and Shigeki Karita and Yifan Ding and Kohei Yatabe and Nobuyuki Morioka and Yu Zhang and Wei Han and Ankur Bapna and Michiel Bacchiani},
      booktitle=WASPAA,
      year={2023},
      eprint={2303.01664},
      archivePrefix={arXiv},
      primaryClass={cs.SD},
      url={https://arxiv.org/abs/2303.01664}, 
}

@inproceedings{Byun_2023,
   title={An Empirical Study on Speech Restoration Guided by Self-Supervised Speech Representation},
   url={http://dx.doi.org/10.1109/ICASSP49357.2023.10095881},
   DOI={10.1109/icassp49357.2023.10095881},
   booktitle=ICASSP,
   publisher={IEEE},
   author={Byun, Jaeuk and Ji, Youna and Chung, Soo-Whan and Choe, Soyeon and Choi, Min-Seok},
   year={2023},
   month=June, pages={1–5} }

@inproceedings{wang2023tfgridnetmakingtimefrequencydomain,
      title={TF-GridNet: Making Time-Frequency Domain Models Great Again for Monaural Speaker Separation}, 
      author={Zhong-Qiu Wang and Samuele Cornell and Shukjae Choi and Younglo Lee and Byeong-Yeol Kim and Shinji Watanabe},
      year={2023},
      booktitle=ICASSP,
      eprint={2209.03952},
      archivePrefix={arXiv},
      primaryClass={cs.SD},
      url={https://arxiv.org/abs/2209.03952}, 
}

@inproceedings{10.1145/1143844.1143891,
author = {Graves, Alex and Fern\'{a}ndez, Santiago and Gomez, Faustino and Schmidhuber, J\"{u}rgen},
title = {Connectionist temporal classification: labelling unsegmented sequence data with recurrent neural networks},
year = {2006},
isbn = {1595933832},
publisher = {Association for Computing Machinery},
address = {New York, NY, USA},
url = {https://doi.org/10.1145/1143844.1143891},
doi = {10.1145/1143844.1143891},
booktitle = ICML,
pages = {369–376},
numpages = {8},
location = {Pittsburgh, Pennsylvania, USA},
series = {ICML '06}
}

@inproceedings{xu2021simpleeffectivezeroshotcrosslingual,
      title={Simple and Effective Zero-shot Cross-lingual Phoneme Recognition}, 
      author={Qiantong Xu and Alexei Baevski and Michael Auli},
      year={2021},
      booktitle=INTERSPEECH,
      eprint={2109.11680},
      archivePrefix={arXiv},
      primaryClass={cs.CL},
      url={https://arxiv.org/abs/2109.11680}, 
}

@inproceedings{song2022exploringwavlmspeechenhancement,
      title={Exploring WavLM on Speech Enhancement}, 
      author={Hyungchan Song and Sanyuan Chen and Zhuo Chen and Yu Wu and Takuya Yoshioka and Min Tang and Jong Won Shin and Shujie Liu},
      year={2022},
      booktitle=SLT,
      eprint={2211.09988},
      archivePrefix={arXiv},
      primaryClass={eess.AS},
      url={https://arxiv.org/abs/2211.09988}, 
}

@inproceedings{xu2019understandingimprovinglayernormalization,
      title={Understanding and Improving Layer Normalization}, 
      author={Jingjing Xu and Xu Sun and Zhiyuan Zhang and Guangxiang Zhao and Junyang Lin},
      year={2019},
      booktitle=NeurIPS,
      eprint={1911.07013},
      archivePrefix={arXiv},
      primaryClass={cs.LG},
      url={https://arxiv.org/abs/1911.07013}, 
}

@misc{mao2017squaresgenerativeadversarialnetworks,
      title={Least Squares Generative Adversarial Networks}, 
      author={Xudong Mao and Qing Li and Haoran Xie and Raymond Y. K. Lau and Zhen Wang and Stephen Paul Smolley},
      year={2017},
      booktitle=ECCV,
      eprint={1611.04076},
      archivePrefix={arXiv},
      primaryClass={cs.CV},
      url={https://arxiv.org/abs/1611.04076}, 
}

@inproceedings{7539284,
  author={Jensen, Jesper and Taal, Cees H.},
  title={An Algorithm for Predicting the Intelligibility of Speech Masked by Modulated Noise Maskers}, 
  year={2016},
  volume={24},
  number={11},
  pages={2009-2022},
  booktitle=TASLP,
  doi={10.1109/TASLP.2016.2585878}}

@inproceedings{hu2021loralowrankadaptationlarge,
      title={LoRA: Low-Rank Adaptation of Large Language Models}, 
      author={Edward J. Hu and Yelong Shen and Phillip Wallis and Zeyuan Allen-Zhu and Yuanzhi Li and Shean Wang and Lu Wang and Weizhu Chen},
      year={2022},
      booktitle=ICLR,
      eprint={2106.09685},
      archivePrefix={arXiv},
      primaryClass={cs.CL},
      url={https://arxiv.org/abs/2106.09685}, 
}

@inproceedings{koizumi2023librittsrrestoredmultispeakertexttospeech,
      title={LibriTTS-R: A Restored Multi-Speaker Text-to-Speech Corpus}, 
      author={Yuma Koizumi and Heiga Zen and Shigeki Karita and Yifan Ding and Kohei Yatabe and Nobuyuki Morioka and Michiel Bacchiani and Yu Zhang and Wei Han and Ankur Bapna},
      booktitle=INTERSPEECH,
      year={2023},
      eprint={2305.18802},
      archivePrefix={arXiv},
      primaryClass={eess.AS},
      url={https://arxiv.org/abs/2305.18802}, 
}

@inproceedings{Wichern2019WHAM,
    title     = {WHAM!: Extending Speech Separation to Noisy Environments},
    author    = {Wichern, Gordon and Antognini, Joe and Flynn, Michael and Zhu,
                 Licheng Richard and McQuinn, Emmett and Crow,
                 Dwight and Manilow, Ethan and Le Roux, Jonathan},
    booktitle = INTERSPEECH,
    year      = {2019},
    month     = sep
}

@inproceedings{reddy2020interspeech2020deepnoise,
      title={The INTERSPEECH 2020 Deep Noise Suppression Challenge: Datasets, Subjective Testing Framework, and Challenge Results}, 
      author={Chandan K. A. Reddy and Vishak Gopal and Ross Cutler and Ebrahim Beyrami and Roger Cheng and Harishchandra Dubey and Sergiy Matusevych and Robert Aichner and Ashkan Aazami and Sebastian Braun and Puneet Rana and Sriram Srinivasan and Johannes Gehrke},
      year={2020},
      eprint={2005.13981},
      booktitle=INTERSPEECH,
      archivePrefix={arXiv},
      primaryClass={eess.AS},
      url={https://arxiv.org/abs/2005.13981}, 
}

@inproceedings{scheibler2024universalscorebasedspeechenhancement,
      title={Universal Score-based Speech Enhancement with High Content Preservation}, 
      author={Robin Scheibler and Yusuke Fujita and Yuma Shirahata and Tatsuya Komatsu},
      year={2024},
      booktitle=Interspeech,
      eprint={2406.12194},
      archivePrefix={arXiv},
      primaryClass={eess.AS},
      url={https://arxiv.org/abs/2406.12194}, 
}

@inproceedings{li2026sensesemanticawarehighfidelityuniversal,
      title={SenSE: Semantic-Aware High-Fidelity Universal Speech Enhancement}, 
      author={Xingchen Li and Hanke Xie and Ziqian Wang and Zihan Zhang and Longshuai Xiao and Shuai Wang and Lei Xie},
      year={2026},
      booktitle=ICME,
      eprint={2509.24708},
      archivePrefix={arXiv},
      primaryClass={eess.AS},
      url={https://arxiv.org/abs/2509.24708}, 
}

@inproceedings{rong2026stupaselowhallucinationstudioqualitygenerative,
      title={StuPASE: Towards Low-Hallucination Studio-Quality Generative Speech Enhancement}, 
      author={Xiaobin Rong and Jun Gao and Zheng Wang and Mansur Yesilbursa and Kamil Wojcicki and Jing Lu},
      year={2026},
      booktitle=Interspeech,
      eprint={2603.09234},
      archivePrefix={arXiv},
      primaryClass={eess.AS},
      url={https://arxiv.org/abs/2603.09234}, 
}

@inproceedings{yao2025gensegenerativespeechenhancement,
      title={GenSE: Generative Speech Enhancement via Language Models using Hierarchical Modeling}, 
      author={Jixun Yao and Hexin Liu and Chen Chen and Yuchen Hu and EngSiong Chng and Lei Xie},
      year={2025},
      booktitle=ICLR,
      eprint={2502.02942},
      archivePrefix={arXiv},
      primaryClass={eess.AS},
      url={https://arxiv.org/abs/2502.02942}, 
}

@inproceedings{liu2024jointsemanticknowledgedistillation,
      title={Joint Semantic Knowledge Distillation and Masked Acoustic Modeling for Full-band Speech Restoration with Improved Intelligibility}, 
      author={Xiaoyu Liu and Xu Li and Joan Serrà and Santiago Pascual},
      booktitle=ICASSP,
      year={2025},
      eprint={2409.09357},
      archivePrefix={arXiv},
      primaryClass={cs.SD},
      url={https://arxiv.org/abs/2409.09357}, 
}

@inproceedings{babaev2024finallyfastuniversalspeech,
      title={FINALLY: fast and universal speech enhancement with studio-like quality}, 
      author={Nicholas Babaev and Kirill Tamogashev and Azat Saginbaev and Ivan Shchekotov and Hanbin Bae and Hosang Sung and WonJun Lee and Hoon-Young Cho and Pavel Andreev},
      year={2024},
      booktitle=NeurIPS,
      eprint={2410.05920},
      archivePrefix={arXiv},
      primaryClass={cs.SD},
      url={https://arxiv.org/abs/2410.05920}, 
}

@inproceedings{saeki2022utmosutokyosarulabvoicemoschallenge,
      title={UTMOS: UTokyo-SaruLab System for VoiceMOS Challenge 2022}, 
      author={Takaaki Saeki and Detai Xin and Wataru Nakata and Tomoki Koriyama and Shinnosuke Takamichi and Hiroshi Saruwatari},
      year={2022},
      eprint={2204.02152},
      archivePrefix={arXiv},
      primaryClass={cs.SD},
      url={https://arxiv.org/abs/2204.02152}, 
}

@inproceedings{rong2025paseleveragingphonologicalprior,
      title={PASE: Leveraging the Phonological Prior of WavLM for Low-Hallucination Generative Speech Enhancement}, 
      author={Xiaobin Rong and Qinwen Hu and Mansur Yesilbursa and Kamil Wojcicki and Jing Lu},
      year={2026},
      booktitle=AAAI,
      eprint={2511.13300},
      archivePrefix={arXiv},
      primaryClass={eess.AS},
      url={https://arxiv.org/abs/2511.13300}, 
}

@inproceedings{lee2026semambageneralspeechrestoration,
      title={SEMamba++: A General Speech Restoration Framework Leveraging Global, Local, and Periodic Spectral Patterns}, 
      author={Yongjoon Lee and Jung-Woo Choi},
      year={2026},
      eprint={2603.11669},
      booktitle=INTERSPEECH,
      archivePrefix={arXiv},
      primaryClass={eess.AS},
      url={https://arxiv.org/abs/2603.11669}, 
}

@inproceedings{nakata2026sidonfastrobustopensource,
      title={Sidon: Fast and Robust Open-Source Multilingual Speech Restoration for Large-scale Dataset Cleansing}, 
      author={Wataru Nakata and Yuki Saito and Yota Ueda and Hiroshi Saruwatari},
      year={2026},
      booktitle=ICASSP,
      eprint={2509.17052},
      archivePrefix={arXiv},
      primaryClass={cs.SD},
      url={https://arxiv.org/abs/2509.17052}, 
}

@INPROCEEDINGS{10363040,
  author={Pirklbauer, Jan and Sach, Marvin and Fluyt, Kristoff and Tirry, Wouter and Wardah, Wafaa and Moeller, Sebastian and Fingscheidt, Tim},
  booktitle={Speech Communication; 15th ITG Conference}, 
  title={Evaluation Metrics for Generative Speech Enhancement Methods: Issues and Perspectives}, 
  year={2023},
  volume={},
  number={},
  pages={265-269},
  doi={10.30420/456164052}}

@INPROCEEDINGS{941023,
  author={Rix, A.W. and Beerends, J.G. and Hollier, M.P. and Hekstra, A.P.},
  booktitle={2001 IEEE International Conference on Acoustics, Speech, and Signal Processing. Proceedings (Cat. No.01CH37221)}, 
  title={Perceptual evaluation of speech quality (PESQ)-a new method for speech quality assessment of telephone networks and codecs}, 
  year={2001},
  volume={2},
  booktitle=ICASSP,
  number={},
  pages={749-752 vol.2},
  doi={10.1109/ICASSP.2001.941023}}

@article{Zhang_2025,
   title={AnyEnhance: A Unified Generative Model With Prompt-Guidance and Self-Critic for Voice Enhancement},
   volume={33},
   ISSN={2998-4173},
   url={http://dx.doi.org/10.1109/TASLPRO.2025.3587393},
   DOI={10.1109/taslpro.2025.3587393},
   journal={IEEE Transactions on Audio, Speech and Language Processing},
   publisher={Institute of Electrical and Electronics Engineers (IEEE)},
   author={Zhang, Junan and Yang, Jing and Fang, Zihao and Wang, Yuancheng and Zhang, Zehua and Wang, Zhuo and Fan, Fan and Wu, Zhizheng},
   year={2025},
   pages={3085–3098} }

@inproceedings{Liu_2022,
   title={VoiceFixer: A Unified Framework for High-Fidelity Speech Restoration},
   url={http://dx.doi.org/10.21437/Interspeech.2022-11026},
   DOI={10.21437/interspeech.2022-11026},
   booktitle=INTERSPEECH,
   publisher={ISCA},
   author={Liu, Haohe and Liu, Xubo and Kong, Qiuqiang and Tian, Qiao and Zhao, Yan and Wang, DeLiang and Huang, Chuanzeng and Wang, Yuxuan},
   year={2022},
   month=sep, collection={interspeech_2022} }

@article{vanDerMaaten2008,
  author = {van der Maaten, Laurens and Hinton, Geoffrey},
  journal = {Journal of Machine Learning Research},
  pages = {2579--2605},
  title = {Visualizing Data using {t-SNE} },
  url = {http://www.jmlr.org/papers/v9/vandermaaten08a.html},
  volume = 9,
  year = 2008
}

@inproceedings{baevski2020wav2vec20frameworkselfsupervised,
      title={wav2vec 2.0: A Framework for Self-Supervised Learning of Speech Representations}, 
      author={Alexei Baevski and Henry Zhou and Abdelrahman Mohamed and Michael Auli},
      year={2020},
      booktitle=NeurIPS,
      eprint={2006.11477},
      archivePrefix={arXiv},
      primaryClass={cs.CL},
      url={https://arxiv.org/abs/2006.11477}, 
}

@inproceedings{yang2026spearunifiedsslframework,
      title={SPEAR: A Unified SSL Framework for Learning Speech and Audio Representations}, 
      author={Xiaoyu Yang and Yifan Yang and Zengrui Jin and Ziyun Cui and Wen Wu and Baoxiang Li and Chao Zhang and Phil Woodland},
      year={2026},
      booktitle=ICML,
      eprint={2510.25955},
      archivePrefix={arXiv},
      primaryClass={eess.AS},
      url={https://arxiv.org/abs/2510.25955}, 
}

@inproceedings{reddy2021dnsmosnonintrusiveperceptualobjective,
      title={DNSMOS: A Non-Intrusive Perceptual Objective Speech Quality metric to evaluate Noise Suppressors}, 
      author={Chandan K A Reddy and Vishak Gopal and Ross Cutler},
      year={2021},
      booktitle=ICASSP,
      eprint={2010.15258},
      archivePrefix={arXiv},
      primaryClass={cs.SD},
      url={https://arxiv.org/abs/2010.15258}, 
}

@article{spearman04,
  author = {Spearman, C.},
  journal = {American Journal of Psychology},
  pages = {88--103},
  title = {The Proof and Measurement of Association Between Two Things},
  volume = 15,
  year = 1904
}

@inproceedings{ragano2025scoreqspeechqualityassessment,
      title={SCOREQ: Speech Quality Assessment with Contrastive Regression}, 
      author={Alessandro Ragano and Jan Skoglund and Andrew Hines},
      year={2025},
      booktitle=NeurIPS,
      eprint={2410.06675},
      archivePrefix={arXiv},
      primaryClass={cs.SD},
      url={https://arxiv.org/abs/2410.06675}, 
}

@inproceedings{fu2024selfsupervisedspeechqualityestimation,
      title={Self-Supervised Speech Quality Estimation and Enhancement Using Only Clean Speech}, 
      author={Szu-Wei Fu and Kuo-Hsuan Hung and Yu Tsao and Yu-Chiang Frank Wang},
      year={2024},
      booktitle=ICLR,
      eprint={2402.16321},
      archivePrefix={arXiv},
      primaryClass={cs.SD},
      url={https://arxiv.org/abs/2402.16321}, 
}

@inproceedings{URGENT-Zhang2024,
  title={{URGENT} Challenge: Universality, Robustness, and Generalizability For Speech Enhancement},
  author={Zhang, Wangyou and Scheibler, Robin and Saijo, Kohei and Cornell, Samuele and Li, Chenda and Ni, Zhaoheng and Pirklbauer, Jan and Sach, Marvin and Watanabe, Shinji and Fingscheidt, Tim and Qian, Yanmin},
  booktitle=INTERSPEECH,
  pages={4868--4872},
  year={2024},
}

@article{vctk,
      title={CSTR VCTK Corpus: English Multi-speaker Corpus for CSTR Voice Cloning Toolkit}, 
      author={Yamagishi, Junichi and Veaux, Christopher and MacDonald, Kirsten},
      year={2019},
      archivePrefix={arXiv},
      primaryClass={eess.AS},
      url={https://doi.org/10.7488/ds/2645}, 
}

@article{prawda2022calibrating,
  title={Calibrating the Sabine and Eyring formulas},
  author={Prawda, Karolina and Schlecht, Sebastian J and V{\"a}lim{\"a}ki, Vesa},
  journal={The Journal of the Acoustical Society of America},
  volume={152},
  number={2},
  pages={1158--1169},
  year={2022},
  publisher={AIP Publishing}
}

@inproceedings{hsu2021hubertselfsupervisedspeechrepresentation,
      title={HuBERT: Self-Supervised Speech Representation Learning by Masked Prediction of Hidden Units}, 
      author={Wei-Ning Hsu and Benjamin Bolte and Yao-Hung Hubert Tsai and Kushal Lakhotia and Ruslan Salakhutdinov and Abdelrahman Mohamed},
      year={2021},
      booktitle=TASLP,
      eprint={2106.07447},
      archivePrefix={arXiv},
      primaryClass={cs.CL},
      url={https://arxiv.org/abs/2106.07447}, 
}

@article{Chen_2022,
   title={WavLM: Large-Scale Self-Supervised Pre-Training for Full Stack Speech Processing},
   volume={16},
   ISSN={1941-0484},
   url={http://dx.doi.org/10.1109/JSTSP.2022.3188113},
   DOI={10.1109/jstsp.2022.3188113},
   number={6},
   journal={IEEE Journal of Selected Topics in Signal Processing},
   publisher={Institute of Electrical and Electronics Engineers (IEEE)},
   author={Chen, Sanyuan and Wang, Chengyi and Chen, Zhengyang and Wu, Yu and Liu, Shujie and Chen, Zhuo and Li, Jinyu and Kanda, Naoyuki and Yoshioka, Takuya and Xiao, Xiong and Wu, Jian and Zhou, Long and Ren, Shuo and Qian, Yanmin and Qian, Yao and Wu, Jian and Zeng, Michael and Yu, Xiangzhan and Wei, Furu},
   year={2022},
   month=Oct, pages={1505–1518} }

@inproceedings{ma2024fleursrrestoredmultilingualspeech,
      title={FLEURS-R: A Restored Multilingual Speech Corpus for Generation Tasks}, 
      author={Min Ma and Yuma Koizumi and Shigeki Karita and Heiga Zen and Jason Riesa and Haruko Ishikawa and Michiel Bacchiani},
      year={2024},
      booktitle=INTERSPEECH,
      eprint={2408.06227},
      archivePrefix={arXiv},
      primaryClass={cs.CL},
      url={https://arxiv.org/abs/2408.06227}, 
}

@string{ICASSP = "ICASSP"}
}
\end{document}